\documentclass[12pt]{article}
\usepackage[letterpaper, portrait, margin=0.83in]{geometry}

\usepackage{amsmath}
\usepackage{amssymb}
\usepackage{times}

\usepackage{ragged2e}
\usepackage{lastpage}
\usepackage{sectsty}
\usepackage{cite}

\usepackage{graphicx}
\sectionfont{\fontsize{12}{15}\selectfont}

\usepackage{caption}
\usepackage[labelfont=bf]{caption}

\usepackage{hyperref}
\hypersetup{
    colorlinks=true,
    linkcolor=blue,
    filecolor=blue,      
    urlcolor=blue,
    citecolor=blue,
    bookmarks=true,
}

\usepackage{setspace}

\usepackage{rotating}
\usepackage{adjustbox}
\usepackage{blindtext}

\usepackage{booktabs}
\usepackage[table,xcdraw]{xcolor}
\usepackage{multirow}
\usepackage{rotating,tabularx}

\begin{document}

\center{\large\textbf{Time-Response Functions of Mechanical Networks with Inerters and Causality}}\bigskip
\center{Nicos Makris}\footnote{Professor, Email: \href{mailto:nicos.makris@ucf.edu}{Nicos.Makris@ucf.edu}}
\center{\footnotesize Department of Civil Engineering, University of Central Florida\\\footnotesize Orlando, Florida, USA}
\bigskip
\hfill
\FlushLeft
\justify
\setlength{\parindent}{0cm} 

\center{\textbf{Abstract}}\justify
\footnotesize
This paper derives the causal time-response functions of three-parameter mechanical networks that have been reported in the literature and involve the inerter--a two-node element in which the force-output is proportional to the relative acceleration of its end-nodes. This two-terminal device is the mechanical analogue of the capacitor in a force-current/velocity-voltage analogy. The paper shows that all frequency-response functions that exhibit singularities along the real frequency axis need to be enhanced with the addition of a Dirac delta function or with its derivative depending on the strength of the singularity. In this way the real and imaginary parts of the enhanced frequency response functions are Hilbert pairs; therefore, yielding a causal time-response function in the time domain. The integral representation of the output signals presented in this paper offers an attractive computational alternative given that the constitutive equations of the three-parameter networks examined herein involve the third derivative of the nodal displacement which may challenge the numerical accuracy of a state-space formulation when the input signal is only available in digital form as in the case of recorded seismic accelerograms.

\normalsize
\justify
\section*{Introduction}\setlength\parindent{0pt}
The force-current; and therefore, velocity-voltage analogy between mechanical and electrical networks \cite{1} respects the in-series and in-parallel configuration of connections, so that equivalent mechanical and electrical networks are expressed by similar diagrams. According to the force-current/velocity-voltage analogy the elastic spring corresponds to the inductor and the linear dashpot corresponds to the resistor. In an effort to lift the constraint that a lumped mass element in a mechanical network has always one of its end-nodes (terminals) connected to the ground, Smith \cite{2} proposed a linear mechanical element that he coined “the inerter” in which the output force is proportional only to the relative acceleration between its end-nodes. Accordingly, the inerter is the precise mechanical analogue of the capacitor. For instance, the driving spinning-top shown in Fig. 1 is a physical realization of the inerter given that the driving force is only proportional to the relative acceleration between terminals 1 and 2. The constant of proportionality of the inerter is coined the "inertance"=$M_R$ [2] and has units of mass $[M]$. The unique characteristic of the inerter is that it has an appreciable inertial mass as oppose to a marginal gravitational mass. Accordingly, if $F_1$, $u_1$ and $F_2$, $u_2$ are the forces and displacements at the end-nodes of the inerter with inertance $M_R$, its constitutive relation is defined as:

\begin{equation}
\left\{ \begin{array}{l}
{F_1}(t)\\
{F_2}(t)
\end{array} \right\} = \left[ {\begin{array}{*{20}{c}}
{{M_R}}&{ - {M_R}}\\
{ - {M_R}}&{{M_R}}
\end{array}} \right]\left\{ \begin{array}{l}
{{\ddot u}_1}(t)\\
{{\ddot u}_2}(t)
\end{array} \right\}
\end{equation}\\
In Eq. (1), the force $F_1(t)=-F_2(t)=M_R(\ddot{u}_1(t)-\ddot{u}_2(t))$ is the through variable of the inerter; whereas, the absolute displacements $u_1$ (respectively $\ddot{u}_1$) and $u_2$ (respectively $\ddot{u}_2$) are the across variables. Smith and his coworkers developed and tested both a rack-and-pinion inerter and a ball-screw inerter \cite{3,4}. Upon its conceptual development and experimental validation, the inerter was implemented to control the suspension vibrations of racing cars under the name of J-damper \cite{5,6}. About the same time a two-terminal flywheel was proposed for the suppression of vehicle vibrations \cite{7}.\\

\begin{figure}[t]
\centering
  \includegraphics[scale=0.75]{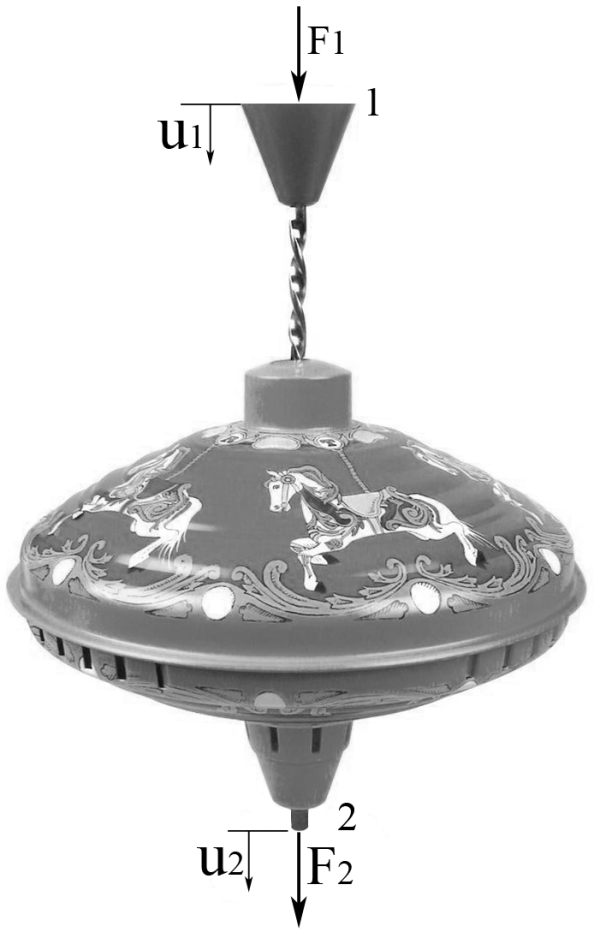}
\caption{A physical realization of the inerter which is the mechanical analogue of the capacitor in a force-current/velocity-voltage analogy.}
\label{fig:1}
\end{figure}

In parallel with the aforementioned developments in vehicle mechanics and dynamics, during the last decade a growing number of publications have proposed the use of rotational inertia dampers for the vibration control and seismic protection of civil structures. For instance, \cite{8} proposed a rotational inertia damper in association with a toggle bracing for vibration control of building structures. The proposed rotational inertia damper consists of a cylindrical mass that is driven by a ball screw and rotates within a chamber that contains some viscous fluid. In this way the vibration reduction originates partly from the difficulty to mobilize the rotational inertia of the rotating mass and partly from the difficulty to shear of the viscous fluid that surrounds the rotating mass. The use of inerters to improve the performance of seismic isolated buildings has been proposed in \cite{9}; while, \cite{10} examined the dynamic response of a single-degree-of-freedom (SDOF) structure equipped with a rotational damper that is very similar to the rotational inertia damper initially proposed in \cite{8}. The main difference is that, in the configuration proposed in \cite{10}, an additional flywheel is appended to accentuate the rotational inertia effect of the proposed vibration control device. About the same time, \cite{11} examined the response of SDOF and multi-degree-of-freedom (MDOF) structures equipped with supplemental rotational inertia that is offered from a ballscrew type device that sets in motion a rotating flywheel. Subsequent studies on the response of MDOF structures equipped with supplemental rotational inertia have been presented by \cite{12,13,14} within the context of enhancing the performance of tuned mass dampers. More recently, \cite{15} showed that the seismic protection of structures with supplemental rotational inertia has some unique advantages, particularly in suppressing the spectral displacement of long period structures---a function that is not efficiently achieved with large values of supplemental damping. However, this happens at the expense of transferring appreciable forces at the support of the flywheels (chevron frames for buildings or end-abutments for bridges).\\

One of the challenges with the dynamic response analysis of civil structures is that while the inerter, or more complex response-modification mechanical networks that involve inerters, are linear networks, the overall structural system in which they belong may behave nonlinearly. In this case the overall structural response needs to be computed in the time-domain. A time-domain representation of the response modification network is possible either via a state-space formulation; or by computing the basic time-response function of the response-modification network and proceeding by solving a set of integro-differential equations. Given that the state-space formulation of some mechanical networks that contain inerters involve the evaluation of the third derivative of the end-node displacement (derivative of the end-node acceleration, see \cite{15} and equations (2) and (3) of this paper), the alternative of calculating the response-history of the through or across variables of the mechanical network by convolving its basic time-response functions becomes attractive. Accordingly, this paper concentrates on deriving the basic time-response functions of practical mechanical networks reported in the literature which involve inerters.

\section*{Motivation and Problem Statement}

Given that the inerter, as defined with Eq. (1), complements the linear spring and the viscous dashpot as the third elementary response-modification element, this paper examines the time-response functions of the three-parameter inertoviscoelastic "fluid" networks shown in Fig. 2 and the three-parameter inertoviscoelastic "solid" networks shown in Fig. 3. The term "fluid" expresses that the network undergoes an infinite displacement under static loading; whereas the term "solid" expresses that the network sustains a finite displacement under a static load.\\
\begin{figure}[t]
\centering
  \includegraphics[scale=0.4]{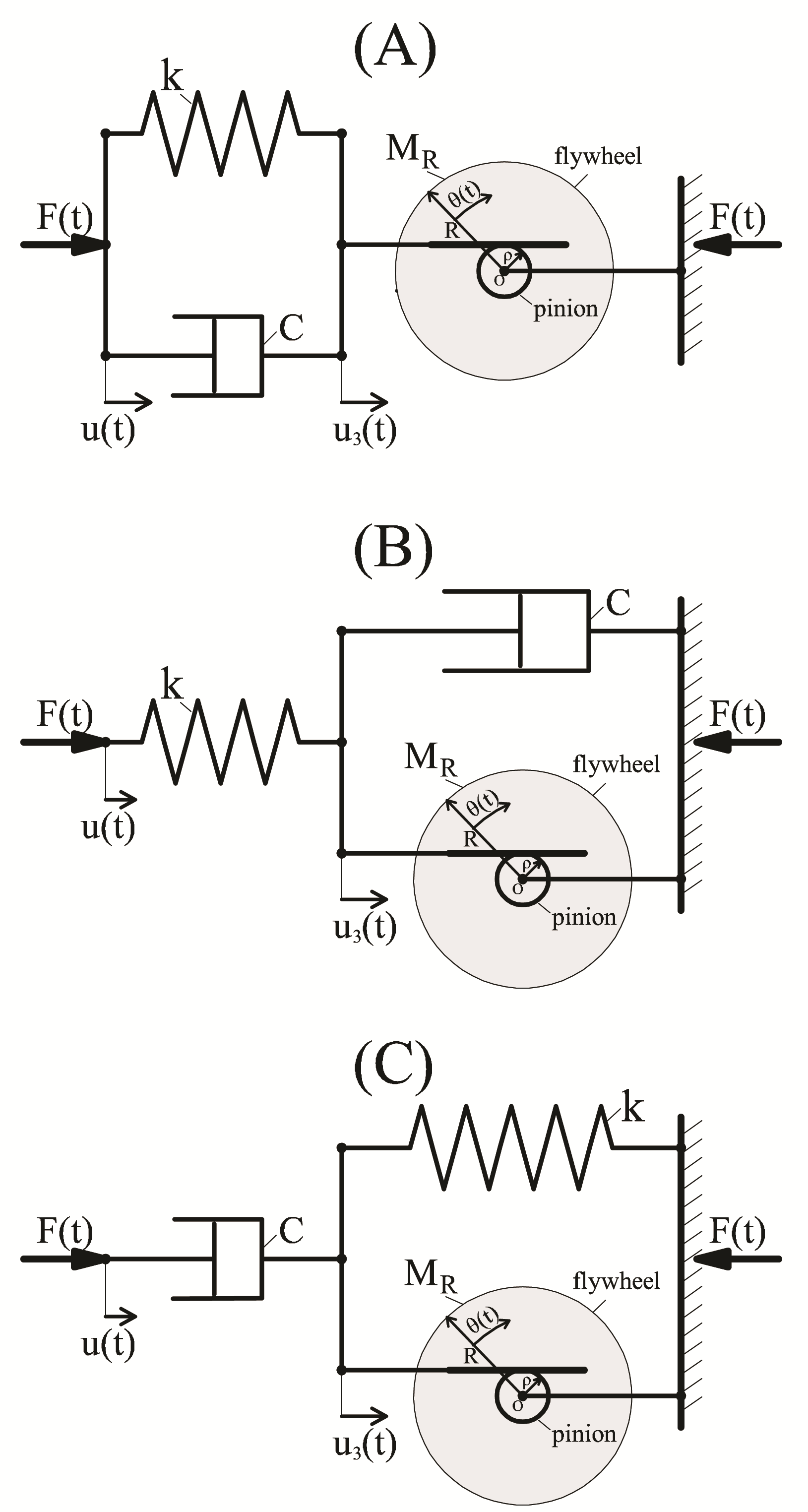}
\caption{The three-parameter inertoviscoelastic fluid A, B and C.}
\label{fig:2}
\end{figure}
\begin{figure}[t]
\centering
  \includegraphics[scale=0.4]{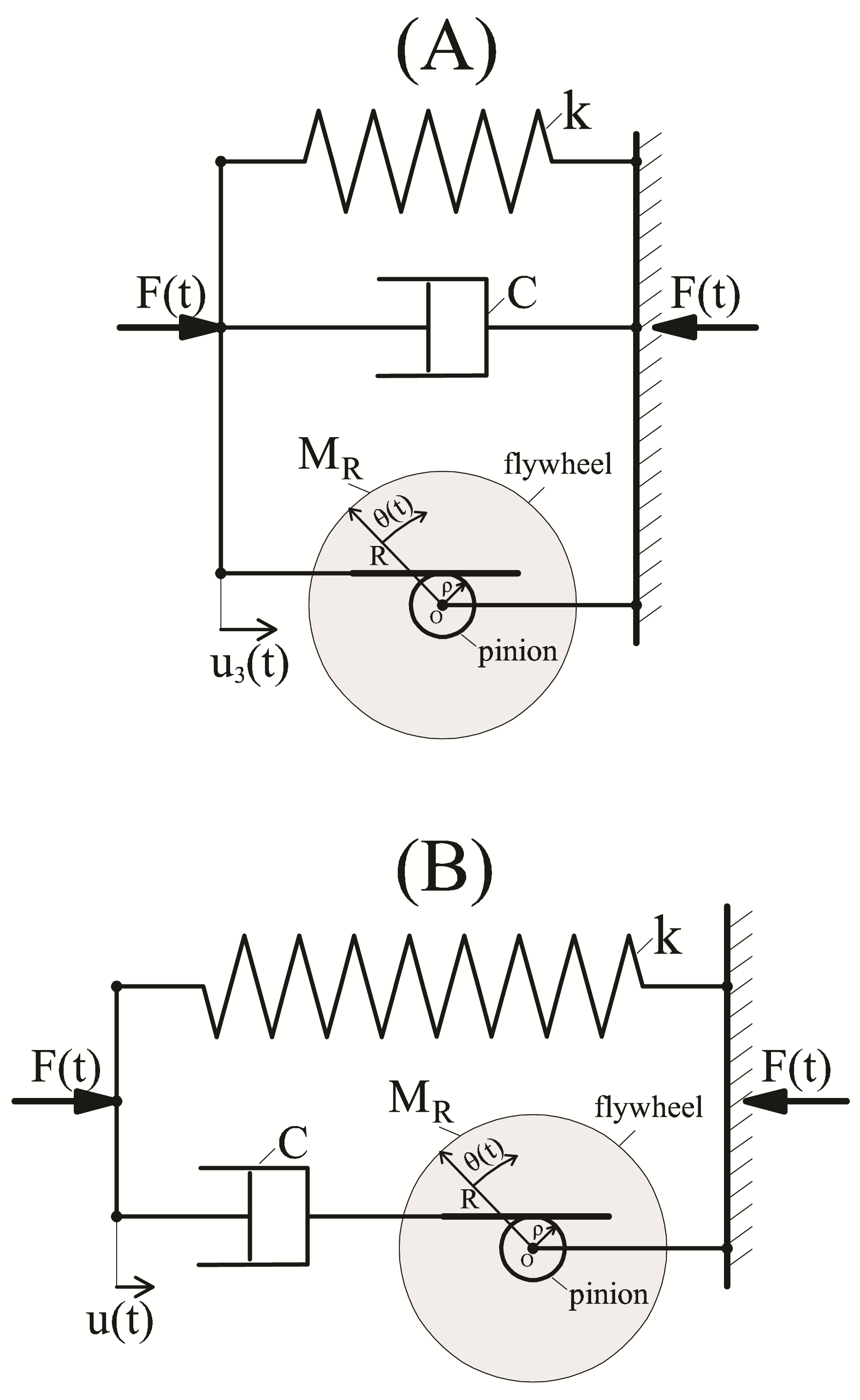}
\caption{The three-parameter inertoviscoelastic solid A and B.}
\label{fig:3}
\end{figure}

Fig. 2 (top) is a spring-dashpot parallel connection (Kelvin-Voight model) that is connected in series with an inerter. This mechanical network, that is coined the inertoviscoelastic fluid A, emerged during the testing of inerters where the spring-dashpot parallel connection served as a mechanical buffer between a prototype inerter and the driving actuator \cite{3}. The effectiveness of the inertoviscoelastic fluid A was subsequently studied extensively by \cite{13} in comparison with the traditional tuned-mass-damper that finds applications in the reduction of building vibrations; whereas, \cite{15} used the same three-parameter model to study the effectiveness of an inerter mounted on a chevron frame for the seismic protection of buildings (or on a bridge abutment for the seismic protection of bridges) with finite stiffness (spring) and damping (dashpot). The constitutive equation of the three-parameter mechanical network shown in Fig. 2 (top) is described in \cite{15},

\begin{equation}
\frac{k}{{{M_R}}}F(t) + \frac{C}{{{M_R}}}\frac{{dF(t)}}{{dt}} + \frac{{{d^2}F(t)}}{{d{t^2}}} = k\frac{{{d^2}u(t)}}{{d{t^2}}} + C\frac{{{d^3}u(t)}}{{d{t^3}}}
\end{equation}
By defining the relaxation time, $\lambda=C/k$ and the rotational frequency $\omega_R=\sqrt{k/M_R}$, Eq. (2) assumes the form:

\begin{equation}
F(t) + \lambda \frac{{dF(t)}}{{dt}} + \frac{1}{{\omega _R^2}}\frac{{{d^2}F(t)}}{{d{t^2}}} = {M_R}\left( {\frac{{{d^2}u(t)}}{{d{t^2}}} + \lambda \frac{{{d^3}u(t)}}{{d{t^3}}}} \right)
\end{equation}\\
The right-hand-side (rhs) of the constitutive equation given by Eqs. (2) or (3) involves the third derivative of the nodal displacement (derivative of the nodal acceleration) and this may challenge the accuracy of the numerically computed response, in particular when the input excitation is only available in digital form as in the case of recorded seismic accelerograms. Part of the motivation of this paper is to bypass this challenge (numerical evaluation of $d^3u(t)/dt^3$) by studying the integral representations of the force, F(t) (through variable), and the displacement, u(t) (end-node variable) appearing in Eqs. (2) or (3).

Upon deriving the time-response functions of the inertoviscoelastic fluid A, the paper proceeds by studying the time-response functions of the inertoviscoelastic fluid B shown in Fig. 2 (center) which is a dashpot-inerter parallel connection (rotational inertia damper) that is connected in series with a spring that approximates the finite stiffness of the mounting connections of a rotational inertia damper \cite{10}. Next, the paper examines the time-response functions of the inertoviscoelastic fluid C shown in Fig. 2 (bottom) which is a spring-inerter parallel connection (inertoelastic solid) that is connected in series with a dashpot. Again, the constitutive equation of the inertoviscoelastic fluid C involves the third derivative of the nodal displacements (derivative of the nodal accelerations) which may challenge the accuracy of the numerical calculation of a state-space formulation. Accordingly, the integral representation of the force and displacement presented in this study offers an attractive alternative.

Finally, the paper examines the time-response functions of the inertoviscoelastic solids A and B shown in Fig. 3. The inertoviscoelastic solid A is a spring-dashpot-inerter parallel connection; whereas, the inertoviscoelastic solid B is a dashpot-inerter in-series arrangement that is connected in parallel with an elastic spring. Inertoviscoelastic solid networks have been proposed for the vibration control of building \cite{9} and vehicles \cite{7}.

\section*{Frequency-and Time-Response Functions}
\label{sec3}
When a combination of springs, dashpots and inerters form a mechanical network, the constitutive equation of the mechanical network is of the form\\
\begin{equation}
\left[ {\sum\limits_{m = 0}^M {{a_m}\frac{{{d^m}}}{{d{t^m}}}} } \right]F(t) = \left[ {\sum\limits_{n = 0}^N {{b_n}\frac{{{d^n}}}{{d{t^n}}}} } \right]u(t)
\end{equation}\\
\noindent where $F(t)$ is the force (through variable) and $u(t)$ is the relative displacement of its end-nodes. In Eq. (4) the coefficients $a_m$ and $b_n$ are restricted to real numbers and the order of differentiation $m$ and $n$ is restricted to integers. The linearity of Eq. (4) permits its transformation in the frequency domain by applying the Fourier transform \\
\begin{equation}
u(\omega ) = H(\omega )F(\omega ) = \left[ {{H_1}(\omega ) + i{H_2}(\omega )} \right]F(\omega )
\end{equation}\\
\noindent where $u(\omega ) = \int\limits_{ - \infty }^\infty  {u(t){e^{ - i\omega t}}dt} $ and $F(\omega ) = \int\limits_{ - \infty }^\infty  {F(t){e^{ - i\omega t}}dt} $ are the Fourier transforms of the relative displacement and force histories respectively; and $H(\omega)$ is the dynamic compliance (dynamic flexibility) of the network:

\begin{equation}
H(\omega ) = \frac{{u(\omega )}}{{F(\omega )}} = \frac{{\sum\limits_{m = 0}^M {{a_m}{{(i\omega )}^m}} }}{{\sum\limits_{n = 0}^N {{b_n}{{(i\omega )}^n}} }}
\end{equation}\\
The dynamic compliance of a mechanical network, $H(\omega)$, as expressed by Eq. (5) is a transfer function that relates a force input to a displacement output. When the dynamic compliance $H(\omega)$ is a proper transfer function, the relative displacement, $u(t)$, in Eq. (4) can be computed in the time domain via the convolution

\begin{equation}
u(t) = \int\limits_{ - \infty }^t {h(t - \tau )F(\tau )d\tau } 
\end{equation}\\
\noindent where $h(t)$ is the "impulse response function" defined as the resulting displacement at time t for an impulsive force input at time $\tau$ ($\tau<t$) and is the inverse Fourier transform of the dynamic compliance

\begin{equation}
h(t) = \frac{1}{{2\pi }}\int\limits_{ - \infty }^\infty  {H(\omega ){e^{i\omega t}}d\omega } 
\end{equation}\\
The mechanical impedance, $Z(\omega)=Z_1(\omega)+iZ_2(\omega)$, is a transfer function which relates a velocity input to a force output

\begin{equation}
F(\omega ) = Z(\omega )v(\omega ) = \left[ {{Z_1}(\omega ) + i{Z_2}(\omega )} \right]v(\omega )
\end{equation}\\
where $v(\omega)=i~\omega~u(\omega)$ is the Fourier transform of the relative velocity time-history. The classical definition of the mechanical impedance as expressed by Eq. (9) (\cite{16,17}, among others) is adopted in this paper given that its corresponding time-response function, known as the relaxation stiffness, $k(t)$ (see Eq. (12)), is a most practical time-response function which can be measured experimentally with a simple relaxation test. Accordingly, for the linear inertoviscoelastic model given by Eq. (4), the mechanical impedance is 

\begin{equation}
Z(\omega ) = \frac{{F(\omega )}}{{v(\omega )}} = \frac{{\sum\limits_{n = 0}^N {{b_n}{{(i\omega )}^n}} }}{{\sum\limits_{m = 0}^M {{a_m}{{(i\omega )}^{m + 1}}} }}
\end{equation}\\
Smith \cite{2} adopts as definition of the mechanical impedance the inverse of the classical definition expressed by Eq. (10) in order to maintain the analogy with electrical engineering where the impedance is the ratio of the voltage across variable (here velocity) to the current though variable (here force).

The force output, F(t) appearing in Eq. (4) can be computed in the time domain with the convolution integral

\begin{equation}
F(t) = \int\limits_{ - \infty }^t {k(t - \tau )\dot u(\tau )d\tau } 
\end{equation}\\
where $k(t)$ is the relaxation stiffness of the mechanical network defined as the resulting force at the present time, $t$, due to a unit step-displacement input at time $\tau$ ($\tau<t$) and is the inverse Fourier transform of the impedance

\begin{equation}
k(t) = \frac{1}{{2\pi }}\int\limits_{ - \infty }^\infty  {Z(\omega ){e^{i\omega t}}d\omega } 
\end{equation}\\
The inverse of the impedance, $Y(\omega)=1/Z(\omega)$, is the admittance; while in the mechanical and structural engineering literature the term "mobility" is used \cite{17}. The admittance (mobility) is a transfer function that relates a force input to a velocity output and when is a proper transfer function, the relative velocity history between the end-nodes of the mechanical network can be computed in the time-domain via the convolution

\begin{equation}
v(t) = \int\limits_{ - \infty }^t {y(t - \tau )F(\tau )d\tau } 
\end{equation}\\
where $y(t)$ is the "impulse velocity response function" defined as the resulting velocity at time t for an impulsive force input at time $\tau$ ($\tau<t$) and is the inverse Fourier transform of the admittance:

\begin{equation}
y(t) = \frac{1}{{2\pi }}\int\limits_{ - \infty }^\infty  {Y(\omega ){e^{i\omega t}}d\omega } 
\end{equation}\\
At negative times ($t<0$), all three time-response functions given by Eqs. (8), (12) and (14) need to be zero in order for the phenomenological model (mechanical network) to be causal. The requirement for a time-response function to be causal in the time domain implies that its corresponding frequency-response function is analytic on the bottom-half complex plane \cite{18,19,20,21,22}. The analyticity condition on a complex function, $Z(\omega)=Z_1(\omega)+iZ_2(\omega)$, relates the real part $Z_1(\omega)$ and the imaginary part $Z_2(\omega)$ with the Hilbert transform \cite{16},\cite{18,19,20}, \cite{23}:

\begin{equation}
{Z_1}(\omega ) =  - \frac{1}{\pi }\int\limits_{ - \infty }^\infty  {\frac{{{{\rm Z}_2}(x)}}{{x - \omega }}dx}~~,~~{Z_2}(\omega ) = \frac{1}{\pi }\int\limits_{ - \infty }^\infty  {\frac{{{{\rm Z}_1}(x)}}{{x - \omega }}dx} 
\end{equation}\\
Prior of computing the time-response functions of the three-parameter mechanical networks shown in Figs. 2 and 3, we first compute the time-response functions of the solitary inerter since its admittance and dynamic compliance exhibit singularities along the real frequency axis and need to be enhanced with either the addition of a Dirac delta function or with its derivative depending on the strength of the singularity.

\section*{Frequency- and Time-Response Functions of The Inerter}
\label{sec4}
The first row of Eq. (1) gives: 

\begin{equation}
F(t) = {M_R}\frac{{{d^2}u(t)}}{{d{t^2}}}
\end{equation}\\
\noindent where  $F(t)=F_1(t)=-F_2(t)$ is the through variable and $u(t)=u_1(t)-u_2(t)$ is the relative displacement of the end-nodes of the inerter. The Fourier transform of Eq. (16) is $F(\omega ) =  - {M_R}{\omega ^2}u(\omega )$; therefore, the compliance of the inerter as defined by (6) is a proper transfer function:

\begin{equation}
H(\omega ) =  - \frac{1}{{{M_R}}}\frac{1}{{{\omega ^2}}}
\end{equation}\\
While the dynamic compliance (dynamic flexibility) of the inerter as expressed by Eq. (17) is a proper transfer function, the inverse Fourier transform of $-1/\omega^2$ is $(t/2).sgn(t)$, where $sgn(t)$ is the signum function. Accordingly, by using the expression of the compliance of the inerter as offered by Eq. (17), the resulting impulse response function as defined by Eq. (8) is $(M_R/2)t.sgn(t)$; which is clearly a non-causal function. In fact the signum function, $sgn(t)$, indicates that there is as much response before the induced impulse force as the response upon the excitation is induced. Two decades ago, this impasse was resolved \cite{21,22} by extending the relation between the analyticity of a transfer function and the causality of the corresponding time-response function to the case where generalized functions are involved \cite{18,19,20}. Given that the compliance of the inerter as expressed by Eq. (17) is a purely real quantity, we are in search of the imaginary Hilbert pair of $-1/\omega^2$.\\

The Hilbert pair of $-1/\omega^2$ is constructed by employing the first of equations (15), together with the property of the derivative of the Dirac delta function \cite{24}:

\begin{equation}
\int\limits_{ - \infty }^\infty  {\frac{{d\delta (t - 0)}}{{dt}}f(t)dt =  - } \int\limits_{ - \infty }^\infty  {\delta (t - 0)\frac{{df(t)}}{{dt}}dt =  - } \frac{{df(0)}}{{dt}}
\end{equation}\\
By letting 
${H_2}(\omega ) = \pi \frac{{d\delta (\omega  - 0)}}{{d\omega }}$, its Hilbert transform gives:

\begin{equation}
{H_1}(\omega ) =  - \frac{1}{\pi }\int\limits_{ - \infty }^\infty  {\pi \frac{{d\delta (x - 0)}}{{dx}}\frac{1}{{x - \omega }}dx} 
\end{equation}\\
and with the change of variables $\xi=x-\omega$, $d\xi=dx$, (19) becomes:

\begin{equation}
\begin{array}{l}
{H_1}(\omega ) =  - \int\limits_{ - \infty }^\infty  {\frac{{d\delta (\xi  - ( - \omega ))}}{{d\xi }}\frac{1}{\xi }d\xi } \\
{\rm{            }} = \int\limits_{ - \infty }^\infty  {\delta (\xi  - ( - \omega ))( - \frac{1}{{{\xi ^2}}})d\xi }  =  - \frac{1}{{{\omega ^2}}}
\end{array}
\end{equation}\\
The result of Eq. (20) indicates that the rhs of Eq. (17) cannot stand alone and has to be accompanied by its imaginary Hilbert pair, $\pi d\delta(\omega-0)/d\omega$. Consequently, the correct expression of the dynamic compliance of the inerter is 

\begin{equation}
H(\omega ) = \frac{1}{{{M_R}}}\left[ { - \frac{1}{{{\omega ^2}}} + i\pi \frac{{d\delta (\omega  - 0)}}{{d\omega }}} \right]
\end{equation}\\
By "manually" appending the imaginary part, $\pi d\delta(\omega-0)/d\omega$, in the rhs of Eq. (17), the inverse Fourier transform of the correct dynamic compliance of the inerter as expressed by Eq. (21) gives:

\begin{equation}
\begin{array}{l}
h(t) = \frac{1}{{2\pi }}\int\limits_{ - \infty }^\infty  {H(\omega ){e^{i\omega t}}d\omega } \\
{\rm{        }} = \frac{1}{{{M_R}}}\frac{1}{{2\pi }}\int\limits_{ - \infty }^\infty  {\left[ { - \frac{1}{{{\omega ^2}}} + i\pi \frac{{d\delta (\omega  - 0)}}{{d\omega }}} \right]} {e^{i\omega t}}d\omega 
\end{array}
\end{equation}\\
By recalling that the Fourier transform of $-1/\omega^2$ is $(t/2)sgn(t)$, Eq. (22) simplifies to

\begin{equation}
h(t) = \frac{1}{{{M_R}}}\left[ {\frac{t}{2}{\mathop{\rm sgn}} (t) + \frac{i}{2}\int\limits_{ - \infty }^\infty  {\frac{{d\delta (\omega  - 0)}}{{d\omega }}} {e^{i\omega t}}d\omega } \right]
\end{equation}\\
and after employing Eq. (18), the second term in the rhs of Eq. (23) gives:

\begin{equation}
\frac{i}{2}\int\limits_{ - \infty }^\infty  {\frac{{d\delta (\omega  - 0)}}{{d\omega }}} {e^{i\omega t}}d\omega  =  - \frac{i}{2}\int\limits_{ - \infty }^\infty  {\delta (\omega  - 0)it} {e^{i\omega t}}d\omega  = \frac{t}{2}
\end{equation}\\
Substitution of the result of Eq. (24) into Eq. (23), gives the causal expression for the impulse response function of the inerter

\begin{equation}
h(t) = \frac{1}{{{M_R}}}\left[ {\frac{t}{2}{\mathop{\rm sgn}} (t) + \frac{t}{2}} \right] = \frac{1}{{{M_R}}}U(t - 0)t
\end{equation}\\
where $U(t-0)$ is the Heaviside unit-step function at the time origin \cite{19,20}. Equation (25) indicates that an impulse force on the inerter creates a causal response that grows linearly with time and is inverse proportional to the inertance, $M_R$.

The impedance of the inerter as defined by Eq. (10) derives directly from Eq. (16) by using that $v(\omega)=i\omega u(\omega)$,

\begin{equation}
Z(\omega ) = i\omega {M_R}
\end{equation}\\
and is an improper transfer function. Accordingly its inverse Fourier transform, that is the relaxation stiffness $k(t)$, as defined by Eq. (12) does not exist in the classical sense. Nevertheless, it can be constructed mathematically with the calculus of generalized functions and more specifically with the property of the derivative of the Dirac delta function given by Eq. (18). By employing Eq. (18), the Fourier transform of $d\delta(t-0)/dt$ is

\begin{equation}
\int\limits_{ - \infty }^\infty  {\frac{{d\delta (t - 0)}}{{dt}}{e^{ - i\omega t}}dt =  - } \int\limits_{ - \infty }^\infty  {\delta (t - 0)( - i\omega ){e^{ - i\omega t}}dt = i\omega } 
\end{equation}\\
Consequently, based on the outcome of Eq. (27), the inverse Fourier transform of the impedance of the inerter given by Eq. (26) is 

\begin{equation}
k(t) = {M_R}\frac{{d\delta (t - 0)}}{{dt}}
\end{equation}\\
Equation (28) indicates that the relaxation stiffness of the inerter exhibits a strong singularity at the time origin given that it is not physically realizable to impose a step displacement to an inerter with finite inertance, $M_R$.

The admittance (mobility) of the inerter is the inverse of its impedance given by Eq. (26):

\begin{equation}
Y(\omega ) = \frac{1}{{{M_R}i\omega }} =  - \frac{1}{{{M_R}}}i\frac{1}{\omega }
\end{equation}\\
Whereas the admittance (mobility) of the inerter as expressed by Eq. (29) is a proper transfer function, the inverse Fourier transform of $-i/\omega$ is $(1/2)sgn(t)$ \cite{16}; where, $sgn(t)$, is the signum function which is clearly a non-causal function. By following the same reasoning described to construct the correct dynamic compliance of the inerter given by Eq. (21) we are in search of the real Hilbert pair of the reciprocal function $-1/\omega$ which is $\pi \delta(\omega-0)$, \cite{19,20,21}, \cite{25}. Accordingly, by appending a Dirac delta function as the real part in Eq. (29), the correct expression of the admittance of the inerter is 

\begin{equation}
Y(\omega ) = \frac{1}{{{M_R}}}\left[ {\pi \delta (\omega  - 0) - i\frac{1}{\omega }} \right]
\end{equation}\\
and the inverse Fourier transform of the correct admittance of the inerter given by Eq. (30) yields

\begin{equation}
y(t) = \frac{1}{{{M_R}}}\left[ {\frac{1}{2} + \frac{1}{2}{\mathop{\rm sgn}} (t)} \right] = \frac{1}{{{M_R}}}U(t - 0)
\end{equation}\\
which is a causal function since $U(t-0)$ is the Heaviside unit-step function at the time origin.\\

The intimate relation between the reciprocal function and the Delta function as expressed by Eq. (30) was first noticed by Dirac \cite{26}. In an effort to make the reciprocal function, $1/x$, well defined in the neighborhood of $x=0$ (in the context of a generalized function), Dirac imposed an extra condition such that the integral of the antisymmetric reciprocal function from $-\epsilon$ to $\epsilon$ $(\forall \epsilon>0)$ to vanish. Accordingly, Dirac \cite{26} demanded that

\begin{equation}
\int_{ - \varepsilon }^\varepsilon  {\frac{1}{x}dx}  = 0
\end{equation}\\
At the same time, if one uses the standard expression from differential calculus,$\frac{{d\ln x}}{{dx}} = \frac{1}{x}$  , the corresponding integral from $-\epsilon$ to $\epsilon$ $(\forall \epsilon>0)$ gives:

\begin{equation}
\int_{ - \varepsilon }^\varepsilon  {\frac{1}{x}dx}  = \int_{ - \varepsilon }^\varepsilon  {\frac{{d\ln x}}{{dx}}} dx = \ln (\varepsilon ) - \ln ( - \varepsilon ) =  - i\pi
\end{equation}\\

Given the contradiction between the results of Eqs. (32) and (33), Dirac \cite{26} explained that as x passes through the zero value (origin of the x-axis) the purely imaginary term, $-i\pi$ in the rhs of Eq. (33) vanishes discontinuously; and therefore, the differentiation of this pure imaginary term yields the term $-i\pi\delta(x-0)$. So the correct expression proposed by Dirac \cite{26} for the derivative of the logarithmic function is:

\begin{equation}
\frac{d}{{dx}}\ln x = \frac{1}{x} - i\pi \delta (x - 0)
\end{equation}\\
which unveils the intimate relation between the reciprocal function, $1/x$ and the Dirac delta function, $\delta(x-0)$. The six basic response functions of the inerter computed in this section are summarized in Table 1.
\begin{table}

  \begin{adjustbox}{addcode={\begin{minipage}{\width}}{\caption{%
      Basic frequency-response functions and their corresponding causal time-response functions of the three-parameter inertoviscoelastic fluid A, B and C.
      }\end{minipage}},rotate=90,center}
      \includegraphics[scale=.9]{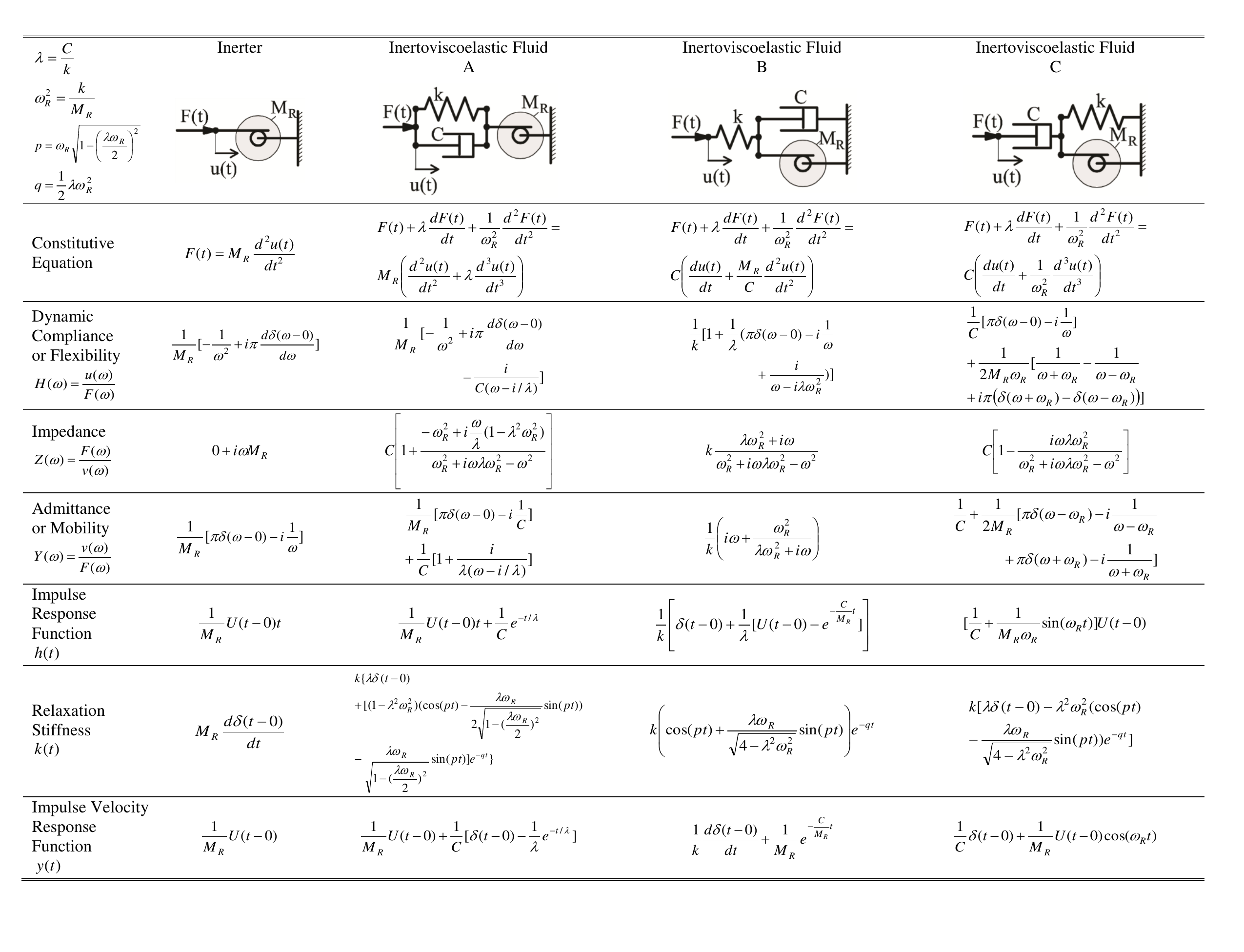}%
  \end{adjustbox}

\label{tab:1}  

\end{table}

\section*{Frequency- and Time-Response Functions of The Three-Parameter Inertoviscoelastic Fluid A}
\label{sec5}
The Fourier transform of the constitutive equation of the inertoviscoelastic fluid A given by Eq. (2) gives

\begin{equation}
(\omega _R^2 + i\omega \lambda \omega _R^2 - {\omega ^2})F(\omega ) =  - {\omega ^2}(k + i\omega C)u(\omega )
\end{equation}\\
Its dynamic compliance, $H(\omega)$, as defined by Eq. (6) is

\begin{equation}
H(\omega ) = \frac{{u(\omega )}}{{F(\omega )}} =  - \frac{{\omega _R^2 + i\omega \lambda \omega _R^2 - {\omega ^2}}}{{{\omega ^2}k(1 + i\omega \lambda )}}
\end{equation}\\
where $\lambda=C/k$ is the relaxation time and $\omega_R=\sqrt{k/M_R}$ is the rotational frequency of the network. Equation (36) indicates that the dynamic compliance of the inertoviscoelastic fluid A has a double pole at $\omega=0$ and a single pole at $\omega=ik/C=i\lambda$. Partial fraction expansion of the rhs of Eq. (36) gives

\begin{equation}
H(\omega ) =  - \frac{1}{{{M_R}}}\frac{1}{{{\omega ^2}}} - \frac{i}{{C(\omega  - i/\lambda )}}
\end{equation}\\
The first term in the rhs of Eq. (37) is the dynamic compliance of the solitary inerter as expressed by Eq. (17); while, the second term is the dynamic compliance of the Kelvin-Voigt model (a spring and a dashpot connected in parallel). Accordingly, the quadratic singularity,$-1/\omega^2$, that is associated with the dynamic compliance of the solitary inerter is enhanced with its imaginary Hilbert companion as shown by Eqs. (19) and (20), and the correct expression for the dynamic compliance of the inertoviscoelastic fluid A is

\begin{equation}
H(\omega ) = \frac{1}{{{M_R}}}\left[ { - \frac{1}{{{\omega ^2}}} + i\pi \frac{{d\delta (\omega  - 0)}}{{d\omega }} - \frac{i}{{C(\omega  - i/\lambda )}}} \right]
\end{equation}\\
Consequently, the dynamic compliance of the inertoviscoelastic fluid A is the superposition of the compliance of the solitary inerter given by Eq. (21) and the compliance of the Kelvin-Voigt model \cite{17}. The inverse Fourier transform of the dynamic compliance as expressed by Eq. (38) gives the causal impulse response function of the inertoviscoelastic fluid A,

\begin{equation}
h(t) = \frac{1}{{{M_R}}}U(t - 0)t + \frac{1}{C}{e^{ - t/\lambda }}
\end{equation}\\
where $U(t-0)$ is again the Heaviside unit-step function at the time origin. \\

In the limiting case of a very soft spring $(k\rightarrow0)$, the relaxation time $\lambda=C/k$ tends to infinity; and therefore, for positive times ($t\geq0$),

\begin{equation}
\mathop {\lim }\limits_{\lambda  \to \infty } {e^{ - t/\lambda }} = U(t - 0)
\end{equation}\\
Consequently, 

\begin{equation}
\mathop {\lim }\limits_{k \to 0} h(t) = \frac{1}{C}\left[ {1 + \frac{C}{{{M_R}}}t} \right]U(t - 0)
\end{equation}\\
which is the impulse response function of a dashpot and an inerter connected in series \cite{27a}.\\

The impedance of the inertoviscoelastic fluid A derives directly from Eq. (35) by using that $v(\omega)=i\omega u(\omega)$ and is given by 

\begin{equation}
Z(\omega ) = \frac{{F(\omega )}}{{v(\omega )}} = \frac{{i\omega k - {\omega ^2}C}}{{\omega _R^2 + i\omega \lambda \omega _R^2 - {\omega ^2}}}
\end{equation}\\
The impedance function given by Eq. (42) is a simple proper transfer function, reaching the constant value, $C$, at the high-frequency limit. By separating the high-frequency limit, $C$, the impedance of the mechanical network shown in Fig. 2 (top) is expressed as

\begin{equation}
Z(\omega ) = C\left[ {1 + \frac{{ - \omega _R^2 + i\frac{\omega }{\lambda }(1 - {\lambda ^2}\omega _R^2)}}{{\omega _R^2 + i\omega \lambda \omega _R^2 - {\omega ^2}}}} \right]
\end{equation}\\
where the frequency-dependent term in the rhs of Eq. (43) is a strictly proper transfer function. The relaxation stiffness, $k(t)$, of the inertoelastic fluid A is the inverse Fourier transform of the impedance given by Eq. (43):

\begin{equation}
k(t) = C\delta (t - 0) + \frac{C}{{2\pi }}\int\limits_{ - \infty }^\infty  {\frac{{\omega _R^2 - i\frac{\omega }{\lambda }(1 - {\lambda ^2}\omega _R^2)}}{{(\omega  - {\omega _1})(\omega  - {\omega _2})}}} {e^{i\omega t}}d\omega 
\end{equation}\\
where $\omega_1$, $\omega_2$ are the poles of the rhs of (43):

\begin{equation}
\begin{array}{l}
{\omega _1} = {\omega _R}\sqrt {1 - {{(\frac{{\lambda {\omega _R}}}{2})}^2}}  + i\frac{\lambda }{2}\omega _R^2 = p + iq\\
\\
{\omega _2} =  - {\omega _R}\sqrt {1 - {{(\frac{{\lambda {\omega _R}}}{2})}^2}}  + i\frac{\lambda }{2}\omega _R^2 =  - p + iq
\end{array}
\end{equation}\\
The inverse Fourier transform of the rhs of Eq. (44) is evaluated with the method of residues and the relaxation stiffness of the three-parameter mechanical network shown in Fig. 2 (top) is

\begin{equation}
\begin{array}{l}
k(t) = C\delta (t - 0) + \\
\frac{1}{p}\left[ {\left( {k - \frac{{{C^2}}}{{{M_R}}}} \right)\left( {p\cos (pt) - q\sin (pt)} \right) - \frac{{Ck}}{{{M_R}}}\sin (pt)} \right]{e^{ - qt}}
\end{array}
\end{equation}\\
where $p = {\omega _R}\sqrt {1 - {{(\lambda {\omega _R}/2)}^2}} $ and $q = \lambda \omega _R^2/2$. Alternatively, by using that $\lambda=C/k$ and $\omega_R^2=k/M_R$, Eq. (46) is expressed as 

\begin{equation}
\begin{array}{l}
k(t) = k\{ \lambda \delta (t - 0)\\
{\rm{           }} + [(1 - {\lambda ^2}\omega _R^2)(\cos (pt) - \frac{{\lambda {\omega _R}}}{{2\sqrt {1 - {{(\frac{{\lambda {\omega _R}}}{2})}^2}} }}\sin (pt))\\
{\rm{           }} - \frac{{\lambda {\omega _R}}}{{\sqrt {1 - {{(\frac{{\lambda {\omega _R}}}{2})}^2}} }}\sin (pt)]{e^{ - qt}}\} 
\end{array}
\end{equation}\\

In the limiting case where the dashpot in the inertoviscoelastic fluid A vanishes, $C=\lambda=q=0$, then $p=\omega_R$ and $e^{-qt}$ tends to $U(t-0)$ for positive times (see Eq. (40)). In this limiting case, Eq. (46) reduces to

\begin{equation}
\mathop {\lim }\limits_{C \to 0} k(t) = kU(t - 0)\cos ({\omega _R}t)
\end{equation}\\
which is the relaxation stiffness of a spring and an inerter connected in series \cite{27a}. Alternatively, when the spring in the inertoviscoelastic fluid A vanishes, $k=1/\lambda=\omega_R=0$, then $p=(i/2)C/M_R$ and $q=(1/2)C/M_R$ and Eq. (47) reduces to

\begin{equation}
\mathop {\lim }\limits_{k \to 0} k(t) = C\left[ {\delta (t - 0) - \frac{C}{{{M_R}}}{e^{ - \frac{C}{{{M_R}}}t}}} \right]
\end{equation}\\
which is the relaxation stiffness of a dashpot and an inerter connected in series \cite{27a}.\\

When the dimensionless quantity, $\lambda \omega_R=2$, then $p=0$, $q=\omega_R$ and the network shown in Fig. 2 (top) becomes critically damped. In this case ($\lambda \omega_R=2$), Eq. (47) assumes the expression

\begin{equation}
\mathop {\lim }\limits_{\lambda {\omega _R} \to 2} k(t) = k\left[ {\lambda \delta (t - 0) + ( - 3 + {\omega _R}t){e^{ - {\omega _R}t}}} \right]
\end{equation}\\
Fig. 4 plots the time history of the non-singular component of the normalized relaxation stiffness, $\frac{{k(t)}}{k} - \lambda \delta (t - 0)$, of the inertoviscoelastic fluid A for four values of $\lambda \omega_R=0.5$, $1$, $1.5$ and $2$.\\

\begin{figure}[t]
\centering
  \includegraphics[scale=0.35]{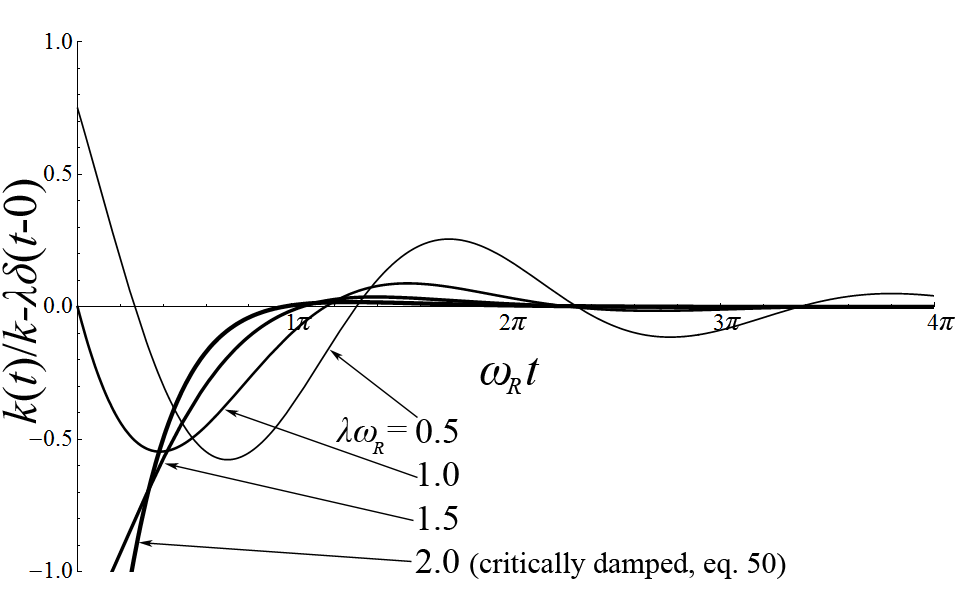}
\caption{Normalized nonsingular term, $\frac{{k(t)}}{k} - \lambda \delta (t - 0)$, of the relaxation stiffness of the inertoviscoelastic fluid A.}
\label{fig:4}
\end{figure}

The admittance (mobility) of the inertoviscoelastic fluid A shown in Fig. 2 (top) is the inverse of its impedance as expressed by Eq. (42); therefore, it is also a simple proper transfer function. By separating its high-frequency limit, $1/C$, the admittance is expressed as

\begin{equation}
Y(\omega ) = \frac{{v(\omega )}}{{F(\omega )}} = \frac{1}{C} + \frac{{i\omega (\lambda \omega _R^2 - 1/\lambda ) + \omega _R^2}}{{\omega k(i - \omega \lambda )}}
\end{equation}\\
where the frequency-dependent term in the rhs of Eq. (51) is a strictly proper transfer function which has a pole at $\omega=0$ and at $\omega=i/\lambda$. Accordingly, partial fraction expansion of the frequency-dependent term gives

\begin{equation}
Y(\omega ) = \frac{1}{C} - \frac{1}{{{M_R}}}i\frac{1}{\omega } + \frac{i}{{\lambda C(\omega  - i/\lambda )}}
\end{equation}\\

The second term in the rhs of Eq. (52) is the admittance of the solitary inerter as expressed by (29). Accordingly, the singularity, $-i/\omega$, that is associated with the admittance of the solitary inerter is enhanced with its real Hilbert companion, $\pi \delta(\omega-0)$, as indicated by Eq. (30), and the correct expression for the admittance of the inertoviscoelastic model A is

\begin{equation}
Y(\omega ) = \frac{1}{{{M_R}}}\left[ {\pi \delta (\omega  - 0) - i\frac{1}{\omega }} \right] + \frac{1}{C}\left[ {1 + \frac{i}{{\lambda (\omega  - i/\lambda )}}} \right]
\end{equation}\\

The second bracket in the rhs of Eq. (53) represent the admittance of the Kelvin-Voight model \cite{17}; therefore, the admittance of the inertoviscoelastic fluid A is the superposition of the admittance of the solitary inerter given by Eq. (30) and the admittance of the Kelvin-Voight model (spring and dashpot in parallel). The inverse Fourier transform of the admittance as expressed by Eq. (53) gives the causal impulse velocity response function of the inertoviscoelastic fluid A

\begin{equation}
y(t) = \frac{1}{{{M_R}}}U(t - 0) + \frac{1}{C}\left[ {\delta (t - 0) - \frac{1}{\lambda }{e^{ - t/\lambda }}} \right]
\end{equation}\\
where $U(t-0)$ is again the Heaviside unit-step function at the time origin. In the limiting case where the spring in the inertoviscoelastic fluid A vanishes, $k=1/\lambda=\omega_R=0$, Eq. (54) reduces to

\begin{equation}
\mathop {\lim }\limits_{k \to 0} y(t) = \frac{1}{C}\delta (t - 0) + \frac{1}{{{M_R}}}U(t - 0)
\end{equation}\\

The six basic response functions of the inertoviscoelastic fluid A shown in Fig. 2 (top) are summarized in Table 1 next to the basic response functions of the solitary inerter.

\section*{Frequency- and Time-Response Functions of The Three-Parameter Inertoviscoelastic Fluid B}
\label{sec6}
The inertoviscoelastic fluid B shown in Fig. 2 (center) is a dashpot-inerter parallel connection (rotational inertia damper) that is connected in series with an elastic spring which in practice may approximate the finite stiffness of the mounting connections of a rotational inertia damper \cite{10}. Given that the force, $F(t)$ is a through variable

\begin{equation}
F(t) = k\left( {u(t) - {u_3}(t)} \right)
\end{equation}\\
while at the same time

\begin{equation}
F(t) = C\frac{{d{u_3}(t)}}{{dt}} + {M_R}\frac{{{d^2}{u_3}(t)}}{{d{t^2}}}
\end{equation}\\
where, $u_3(t)$ is the displacement of the internal node 3. Upon taking the first and second time derivatives of Eq. (56) and substituting the values of the $du_3(t)/dt$ and $d^2u_3(t)/dt^2$, equation (57) yields the constitutive equation of the three-parameter mechanical network shown in Fig. 2 (center)

\begin{equation}
F(t) + \lambda \frac{{dF(t)}}{{dt}} + \frac{1}{{\omega _R^2}}\frac{{{d^2}F(t)}}{{d{t^2}}} = C\left( {\frac{{du(t)}}{{dt}} + \frac{{{M_R}}}{C}\frac{{{d^2}u(t)}}{{d{t^2}}}} \right)
\end{equation}\\
where again $\lambda=C/k$ is the relaxation time and ${\omega _R} = \sqrt {k/{M_R}}$ is the rotational frequency of the network.\\

The Fourier transform of the constitutive equation of the inertoviscoelastic fluid B given by Eq. (58) gives

\begin{equation}
F(\omega )\left( {\omega _R^2 + i\omega \lambda \omega _R^2 - {\omega ^2}} \right) = \omega k\left( {i\lambda \omega _R^2 - \omega } \right)u(\omega )
\end{equation}\\

Its dynamic compliance, $H(\omega)$, as defined by Eq. (5) is 

\begin{equation}
H(\omega ) = \frac{{u(\omega )}}{{F(\omega )}} = \frac{1}{k}\frac{{\omega _R^2 + i\omega \lambda \omega _R^2 - {\omega ^2}}}{{\omega \left( {i\lambda \omega _R^2 - \omega } \right)}}
\end{equation}\\

The dynamic compliance given by Eq. (60) is a simple proper transfer function, reaching the constant, $1/k$, at the high-frequency limit. By separating the high-frequency limit, $1/k$, and upon proceeding with partial fraction expansion of the frequency-dependent term which has simple poles at $\omega=0$ and $\omega=i\lambda \omega_R^2$, the dynamic compliance of the mechanical network shown in Fig. 2 (center) is expressed as  

\begin{equation}
H(\omega ) = \frac{1}{k}\left[ {1 + \frac{1}{\lambda }\left( { - i\frac{1}{\omega } + \frac{i}{{\omega  - i\lambda \omega _R^2}}} \right)} \right]
\end{equation}\\
By following the same reasoning presented when constructing  the correct expression for the admittance of the inerter, the imaginary reciprocal function, $-i/\omega$ appearing in Eq. (61) is enhanced with its real Hilbert companion, $\pi \delta(\omega-0)$ and the correct expression for the dynamic compliance of the inertoviscoelastic fluid B is

\begin{equation}
H(\omega ) = \frac{1}{k}\left[ {1 + \frac{1}{\lambda }\left( {\pi \delta (\omega  - 0) - i\frac{1}{\omega } + \frac{i}{{\omega  - i\lambda \omega _R^2}}} \right)} \right]
\end{equation}\\

The inverse Fourier transform of the dynamic compliance as expressed by Eq. (62) gives the causal impulse response function of the inertoviscoelastic fluid B,

\begin{equation}
h(t) = \frac{1}{k}\left[ {\delta (t - 0) + \frac{1}{\lambda }\left( {U(t - 0) - {e^{ - \frac{C}{{{M_R}}}t}}} \right)} \right]
\end{equation}\\

In the limiting case where the dashpot in the inertoviscoelastic fluid B vanishes, $C=\lambda=0$, the exponential term in the rhs of Eq. (63) is expanded up to the linear term for positive times ($t\geq0$) 

\begin{equation}
\mathop {\lim }\limits_{C \to 0} {e^{ - \frac{C}{{{M_R}}}t}} = \left[ {1 - \frac{C}{{{M_R}}}t + ...} \right]U(t - 0)
\end{equation}\\
where $U(t-0)$ is again the Heaviside unit-step function at the time origin. In this case,

\begin{equation}
\mathop {\lim }\limits_{C \to 0} h(t) = \frac{1}{k}\delta (t - 0) + \frac{1}{{{M_R}}}U(t - 0)t
\end{equation}\\
which is the impulse response function of a spring and an inerter in series \cite{27a}.

The impedance of the inertoviscoelastic fluid B derives directly from Eq. (59) by using that $v(\omega)=i\omega u(\omega)$ and is given by

\begin{equation}
Z(\omega ) = \frac{{F(\omega )}}{{v(\omega )}} = k\frac{{\lambda \omega _R^2 + i\omega }}{{\omega _R^2 + i\omega \lambda \omega _R^2 - {\omega ^2}}}
\end{equation}\\

The impedance function given by Eq. (66) is a strictly proper transfer function and its poles are given by Eq. (45). The inverse Fourier transform of the impedance given by Eq. (66) is the relaxation stiffness of the mechanical network shown in Fig. 2 (center) and is evaluated with the method of residues

\begin{equation}
k(t) = t\left( {\cos (pt) + \frac{{\lambda {\omega _R}}}{{\sqrt {4 - {\lambda ^2}\omega _R^2} }}\sin (pt)} \right){e^{ - qt}}
\end{equation}\\
where again, $p = {\omega _R}\sqrt {1 - {{\left( {\lambda {\omega _R}/2} \right)}^2}} $ and $q = \lambda \omega _R^2/2$. In the limiting case where the dashpot in the inertoviscoelastic fluid B vanishes, $C=\lambda=q=0$, then $p=\omega_R$ and $e^{-qt}$ tends to $U(t-0)$ (see Eq. (40)). In this case Eq. (67) reduces to Eq. (48) which is the relaxation stiffness of a spring and an inerter in series.

When the dimensionless quantity $\lambda \omega_R=2$, then $p=0$, $q=\omega_R$ and the network shown in Fig. 2 (center) becomes critically damped. In this case ($\lambda \omega_R=2$), Eq. (67) assumes the expression:

\begin{equation}
\mathop {\lim }\limits_{\lambda {\omega _R} \to 2} k(t) = k(1 + {\omega _R}t){e^{ - {\omega _R}t}}
\end{equation}\\

Fig. 5 plots the time history of the normalized relaxation stiffness $k(t)/k$ of the inertoviscoelastic fluid B for four values of $\lambda \omega_R=0.5$, $1$, $1.5$ and $2$.
\begin{figure}[t]
  \includegraphics[scale=0.35]{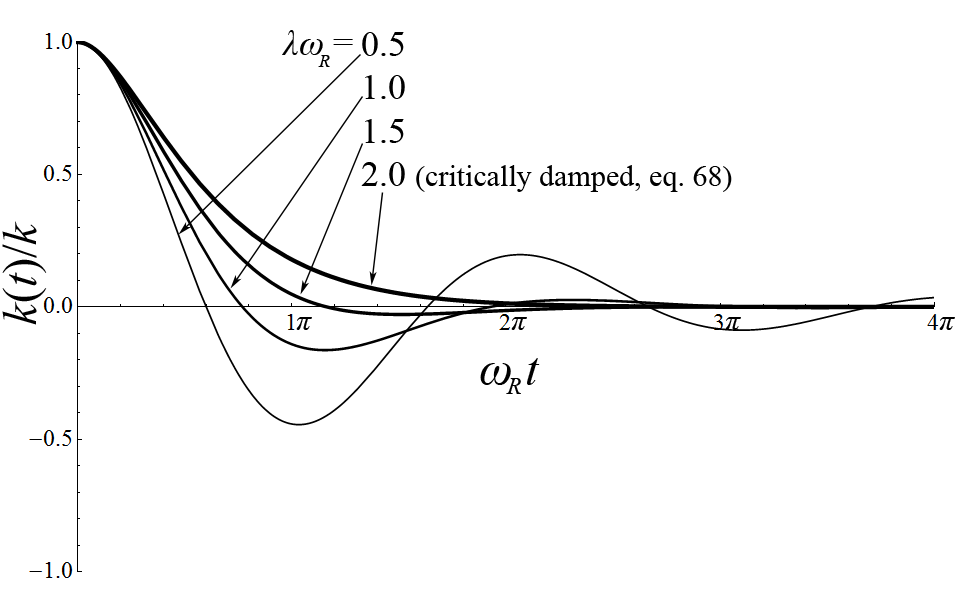}
  \centering
\caption{Normalized relaxation stiffness, $k(t)/k$, of the inertoviscoelastic fluid B. Because of the inerter the kernel $k(t)/k$ exhibits a decaying oscillatory behaviour.}
\label{fig:5}
\end{figure}

The admittance (mobility) of the inertoviscoelastic fluid B shown in Fig. 2 (center) is the inverse of its impedance as expressed by Eq. (66); therefore, it is an improper transfer function. Accordingly, its inverse Fourier transform, that is the impulse velocity response function, $y(t)$, as defined by Eq. (14) does not exist in the classical sense. Nevertheless, it can be constructed mathematically with the calculus of generalized functions. By separating its high-frequency limit, $i\omega/k$, the admittance of the inertoviscoelastic fluid B is expressed as 

\begin{equation}
Y(\omega ) = \frac{{v(\omega )}}{{F(\omega )}} = \frac{1}{k}\left( {i\omega  + \frac{{\omega _R^2}}{{\lambda \omega _R^2 + i\omega }}} \right)
\end{equation}\\

The first term in the rhs of Eq. (69), $i\omega$, is the Fourier transform of $d\delta(t-0)/dt$ as shown by Eq. (27); while the inverse Fourier transform of the second term in the rhs of Eq. (69) is evaluated with the method of residues. Accordingly, the inverse Fourier transform of the admittance given by Eq. (69) gives the impulse velocity response function of the inertoviscoelastic fluid B

\begin{equation}
y(t) = \frac{1}{k}\frac{{d\delta (t - 0)}}{{dt}} + \frac{1}{{{M_R}}}{e^{ - \frac{C}{{{M_R}}}t}}
\end{equation}\\

In the limiting case where the dashpot in the inertoviscoelastic fluid B vanishes, C=0, the exponential term ${e^{ - \frac{C}{{{M_R}}}t}}$ tends to $U(t-0)$ for positive times and Eq. (70) reduces to

\begin{equation}
\mathop {\lim }\limits_{C \to 0} y(t) = \frac{1}{k}\frac{{d\delta (t - 0)}}{{dt}} + \frac{1}{{{M_R}}}U(t - 0)
\end{equation}\\
which is the impulse velocity response function of a spring and an inerter in series \cite{27a}. The six basic response functions of the inertoviscoelastic fluid B shown in Fig. 2 (center) are summarized in Table 1 next to the basic response functions of the inertoviscoelastic fluid A.

\section*{Frequency- and Time-Response Functions of The Three-Parameter Inertoviscoelastic Fluid C}
\label{sec7}

The inertoviscoelastic fluid C shown in Fig. 2 (bottom) is a spring-inerter parallel connection (inertoelastic solid), that is connected in series with a dashpot. Given that the force, F(t) is a through variable,

\begin{equation}
F(t) = C\left( {\frac{{du(t)}}{{dt}} - \frac{{d{u_3}(t)}}{{dt}}} \right)
\end{equation}\\
while at the same time

\begin{equation}
F(t) = k{u_3}(t) + {M_R}\frac{{{d^2}{u_3}(t)}}{{d{t^2}}}
\end{equation}\\
where $u_3(t)$ is the displacement of the internal node 3. Upon taking the first and second derivatives of Eq. (72), Eq. (73) in association with its time-derivative gives

\begin{equation}
F(t) + \lambda \frac{{dF(t)}}{{dt}} + \frac{1}{{\omega _R^2}}\frac{{{d^2}F(t)}}{{d{t^2}}} = C\left( {\frac{{du(t)}}{{dt}} + \frac{1}{{\omega _R^2}}\frac{{{d^3}u(t)}}{{d{t^3}}}} \right)
\end{equation}\\
where again $\lambda=C/k$ is the relaxation time and ${\omega _R} = \sqrt {k/{M_R}} $ is the rotational frequency of the network.

The Fourier transform of the constitutive equation of the inertoviscoelastic fluid C given by Eq. (74) gives

\begin{equation}
F(\omega )\left( {\omega _R^2 + i\omega \lambda \omega _R^2 - {\omega ^2}} \right) = i\omega C\left( {\omega _R^2 - {\omega ^2}} \right)u(\omega )
\end{equation}\\

Its dynamic compliance, $H(\omega)$, as defined by Eq. (5) is 

\begin{equation}
H(\omega ) = \frac{{u(\omega )}}{{F(\omega )}} =  - \frac{{\omega _R^2 + i\omega \lambda \omega _R^2 - {\omega ^2}}}{{i\omega C\left( {{\omega ^2} - \omega _R^2} \right)}}
\end{equation}\\

Equation (76) indicates that the dynamic compliance of the inertoviscoelastic fluid C is a strictly proper transfer function; nevertheless, all its poles $\omega=0$, $\omega=\omega_R$ and $\omega=-\omega_R$ lie on the real axis. Partial fraction expansion of the rhs of Eq. (76) gives

\begin{equation}
H(\omega ) =  - \frac{1}{C}\frac{i}{\omega } - \frac{1}{{2{M_R}}}\frac{1}{{{\omega _R}}}\left[ {\frac{1}{{\omega  - {\omega _R}}} - \frac{1}{{\omega  + {\omega _R}}}} \right]
\end{equation}\\

The first term in the rhs of Eq. (77) is the dynamic compliance of the solitary dashpot \cite{17}; while, the second term with the brackets it the dynamic compliance of a spring and an inerter connected in parallel (the inertoelastic solid). Accordingly, the singularity, $-i/\omega$, that is associated with the dynamic compliance of the solitary dashpot is enhanced with its real Hilbert companion, $\pi \delta(\omega-0)$, as indicated by Eq. (30), while the two singularities $1/(\omega-\omega_R)$ and $1/(\omega+\omega_R)$ appearing within the brackets of the rhs of Eq. (77) are also enhanced with their imaginary Hilbert companions \cite{21,22}, \cite{27a,27}. Following these operations, the correct expression for the dynamic compliance of the inertoviscoelastic fluid C is

\begin{equation}
\begin{array}{l}
H(\omega ) = \frac{1}{C}\left[ {\pi \delta (\omega  - 0) - i\frac{1}{\omega }} \right] + \\
\frac{1}{{2{M_R}{\omega _R}}}\left[ {\frac{1}{{\omega  + {\omega _R}}} - \frac{1}{{\omega  - {\omega _R}}} + i\pi \left( {\delta (\omega  + {\omega _R}) - \delta (\omega  - {\omega _R})} \right)} \right]
\end{array}
\end{equation}\\

The inverse Fourier transform of the dynamic compliance as expressed by Eq. (78) gives the causal impulse response function of the inertoviscoelastic fluid C shown at the bottom of Fig. 2

\begin{equation}
h(t) = \left[ {\frac{1}{C} + \frac{1}{{{M_R}{\omega _R}}}\sin ({\omega _R}t)} \right]U(t - 0)
\end{equation}\\
where $U(t-0)$ is again the Heaviside unit-step function at the time origin.

The impedance of the inertoviscoelastic fluid C derives directly from Eq. (75) by using that $v(\omega)=i\omega u(\omega)$ and is given by

\begin{equation}
Z(\omega ) = \frac{{F(\omega )}}{{v(\omega )}} = C\frac{{\omega _R^2 - {\omega ^2}}}{{\omega _R^2 + i\omega \lambda \omega _R^2 - {\omega ^2}}}
\end{equation}\\

The impedance function given by Eq. (80) is a simple proper transfer function, reaching the constant value, $C$, at the high-frequency limit. By separating the high-frequency limit, $C$, the impedance of the mechanical network shown in Fig. 2 (bottom) is expressed as

\begin{equation}
Z(\omega ) = C\left[ {1 - \frac{{i\omega \lambda \omega _R^2}}{{\omega _R^2 + i\omega \lambda \omega _R^2 - {\omega ^2}}}} \right]
\end{equation}\\

The frequency-dependent term within the brackets of Eq. (81) is a strictly proper transfer function and its poles are given by Eq. (45). The inverse Fourier transform of the impedance function given by Eq. (81) is the relaxation stiffness of the three-parameter mechanical network shown in Fig. 2 (bottom) and is evaluated with the method of residues

\begin{equation}
\begin{array}{l}
k(t) = k[\lambda \delta (t - 0) - {\lambda ^2}\omega _R^2(\cos (pt)\\
{\rm{         }} - \frac{{\lambda {\omega _R}}}{{\sqrt {4 - {\lambda ^2}\omega _R^2} }}\sin (pt)){e^{ - qt}}]
\end{array}
\end{equation}\\
where again, $p = {\omega _R}\sqrt {1 - {{\left( {\lambda {\omega _R}/2} \right)}^2}} $ and $q = \lambda \omega _R^2/2$. In the limiting case where the spring in the inertoviscoelastic fluid C vanishes, $k=1/\lambda=\omega_R=0$, then $p=(i/2)C/M_R$ and $q=(1/2)CM_R$ and Eq. (82) reduces to Eq. (49), which is the relaxation stiffness of a dashpot and an inerter connected in series.

When the dimensionless quantity, $\lambda \omega_R=2$, then $p=0$, $q=\omega_R$ and the network shown in Fig. 2 (bottom) becomes critically damped. In this case ($\lambda \omega_R=2$), Eq. (82) assumes the expression

\begin{equation}
\mathop {\lim }\limits_{\lambda {\omega _R} \to 2} k(t) = k\left[ {\lambda \delta (t - 0) - 4(1 - {\omega _R}t){e^{ - {\omega _R}t}}} \right]
\end{equation}\\

Fig. 6 plots the time history of the non-singular component of the normalized relaxation stiffness $\frac{{k(t)}}{k} - \lambda \delta (t - 0)$ of the inertoviscoelastic fluid C for four values of $\lambda \omega_R=0.5$, $1$, $1.5$ and $2$.
\begin{figure}[t]
\centering
  \includegraphics[scale=0.35]{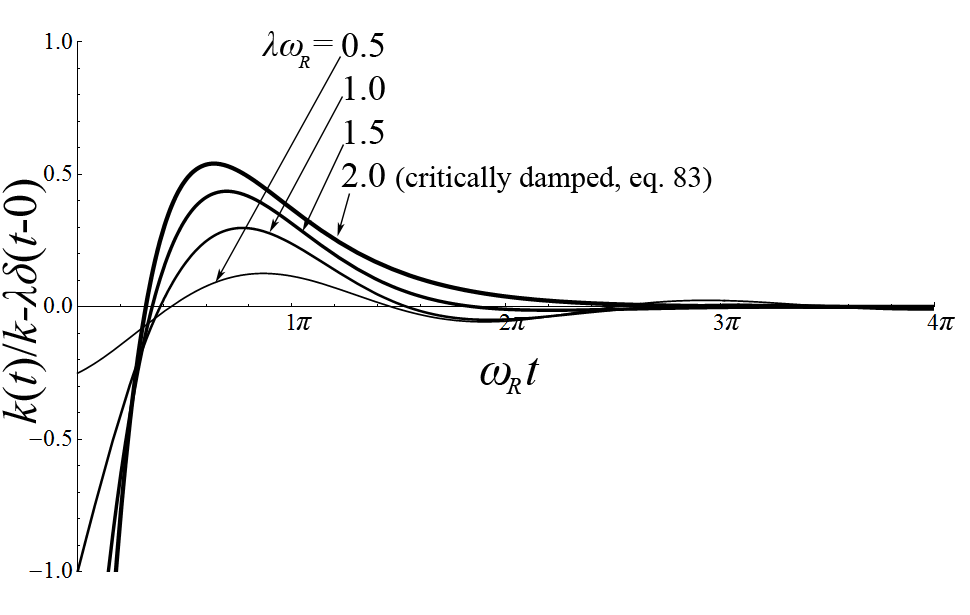}
\caption{Normalized nonsingular term, $\frac{{k(t)}}{k} - \lambda \delta (t - 0)$, of the relaxation stiffness of the inertoviscoelastic fluid C.}
\label{fig:6}
\end{figure}

The admittance (mobility) of the inertoviscoelastic fluid C shown in Fig. 2 (bottom) is the inverse of its impedance as expressed by Eq.  (80); therefore, it is a simple proper transfer function. By separating its high-frequency limit, $1/C$, the admittance of the inertoviscoelastic fluid C is expressed as

\begin{equation}
Y(\omega ) = \frac{{v(\omega )}}{{F(\omega )}} = \frac{1}{C} - i\frac{1}{{{M_R}}}\frac{\omega }{{(\omega  - {\omega _R})(\omega  + {\omega _R})}}
\end{equation}\\
where the real constant term, $1/C$, is the admittance of the solitary dashpot and the imaginary frequency-dependent term in the rhs of Eq. (84) is the admittance of a spring and an inerter connected in parallel. Because the singularities of the rhs term, $ \omega=\omega_R$ and $ \omega=-\omega_R$ lie on the real frequency axis they are enhanced with their real Hilbert companions \cite{21,22}. Following this operation, the correct expression for the admittance of the inertoviscoelastic fluid C shown at the bottom of Fig. 2 is

\begin{equation}
\begin{array}{r}
Y(\omega ) = \frac{1}{C} + \frac{1}{{2{M_R}}}[\pi \delta (\omega  - {\omega _R}) - i\frac{1}{{\omega  - {\omega _R}}}\\
 + \pi \delta (\omega  + {\omega _R}) - i\frac{1}{{\omega  + {\omega _R}}}]
\end{array}
\end{equation}\\

The inverse Fourier transform of the admittance (mobility) as expressed by Eq. (85) gives the causal impulse velocity response function of the inertoviscoelastic fluid C shown at the bottom of Fig. 2

\begin{equation}
y(t) = \frac{1}{C}\delta (t - 0) + \frac{1}{{{M_R}}}U(t - 0)\cos ({\omega _R}t)
\end{equation}\\
where $U(t-0)$ is again the Heaviside unit-step function at the time origin. The six basic response functions of the inertoviscoelastic fluid C shown in Fig. 2 (bottom) are summarized in Table 1 next to the basic response functions of the inertoviscoelastic fluids A and B.

\section*{Frequency- and Time-Response Functions of The Three-Parameter Inertoviscoelastic Solid A}
\label{sec8}

The inertoviscoelastic solid A shown in Fig. 3 (top) consists of a spring, a dashpot and an inerter connected in parallel and has been proposed as a conceptual vibration-control element for buildings \cite{9} and vehicles \cite{7}.

The constitutive equation of the three-parameter mechanical network shown in Fig. 3 (top) is

\begin{equation}
F(t) = ku(t) + C\frac{{du(t)}}{{dt}} + {M_R}\frac{{{d^2}u(t)}}{{d{t^2}}}
\end{equation}\\

Upon applying Fourier transform to Eq. (87), the dynamic compliance, $H(\omega)$ of the inertoviscoelastic solid A is

\begin{equation}
H(\omega ) = \frac{{u(\omega )}}{{F(\omega )}} =  - \frac{1}{{{M_R}}}\frac{1}{{{\omega ^2} - i\omega \lambda \omega _R^2 - \omega _R^2}}
\end{equation}\\
which is a strictly proper transfer function and its poles are given by Eq. (45). The inverse Fourier transform of the dynamic compliance given by Eq. (88) is the impulse response function of the inertoviscoelastic solid A

\begin{equation}
h(t) = \frac{1}{{{M_R}}}\frac{1}{p}\sin (pt){e^{ - qt}}
\end{equation}\\
where again, where $p = {\omega _R}\sqrt {1 - {{(\lambda {\omega _R}/2)}^2}} $ and $q = \lambda \omega _R^2/2$. When the dimensionless quantity, $\lambda \omega_R=2$, then $p=0$, $q=\omega_R$ and the inertoviscoelastic solid A becomes critically damped. In this case ($\lambda \omega_R=2$), Eq. (89) assumes the expression

\begin{equation}
\mathop {\lim }\limits_{\lambda {\omega _R} \to 2} h(t) = \frac{1}{{{M_R}}}t{e^{ - {\omega _R}t}}
\end{equation}\\

The impedance of the inertoviscoelastic solid A derives directly from Eq. (87) by using that $v(\omega)=i\omega u(\omega)$ and is given by

\begin{equation}
Z(\omega ) = \frac{{F(\omega )}}{{v(\omega )}} =  - k\frac{i}{\omega } + C + i\omega {M_R}
\end{equation}\\

The first term in the rhs of Eq. (91) is the impedance of the solitary spring. Accordingly, the singularity, $-i/\omega$, is enhanced with its real Hilbert companion, $\pi \delta(\omega-0)$, as indicated by Eq. (30) and the correct expression for the impedance of the inertoviscoelastic solid A is

\begin{equation}
Z(\omega ) = k\left[ {\pi \delta (\omega  - 0) - i\frac{1}{\omega }} \right] + C + i\omega {M_R}
\end{equation}\\

Equation (92) indicates that the mechanical impedance of the three-parameter network shown in Fig. 3 (top) is the summation of the individual impedances of the solitary spring, dashpot and inerter. Accordingly, the inverse Fourier transform of the impedance as expressed by Eq. (92) gives the causal relaxation stiffness of the inertoviscoelastic solid A:

\begin{equation}
k(t) = kU(t - 0) + C\delta (t - 0) + {M_R}\frac{{d\delta (t - 0)}}{{dt}}
\end{equation}\\

The admittance (mobility) of the inertoviscoelastic solid A derives directly from Eq. (88) by using that $v(\omega)=i\omega u(\omega)$

\begin{equation}
Y(\omega ) = \frac{{v(\omega )}}{{F(\omega )}} =  - \frac{1}{{{M_R}}}\frac{{i\omega }}{{{\omega ^2} - i\omega \lambda \omega _R^2 - \omega _R^2}}
\end{equation}\\
which is a strictly proper transfer function and its poles are again given by Eq. (45). The inverse Fourier transform of the admittance gives the impulse velocity response function of the inertoviscoelastic solid A

\begin{equation}
y(t) = \frac{1}{{{M_R}}}\left[ {\cos (pt) - \frac{q}{p}\sin (pt)} \right]{e^{ - qt}}
\end{equation}\\
where again, $p = {\omega _R}\sqrt {1 - {{\left( {\lambda {\omega _R}/2} \right)}^2}} $ and  $q = \lambda \omega _R^2/2$. 

When the dimensionless quantity, $\lambda \omega_R=2$, then $p=0$, $q=\omega_R$ and the three-parameter mechanical network shown in Fig. 3 (top) becomes critically damped. In this case ($\lambda \omega_R=2$), Eq. (94) assumes the expression

\begin{equation}
\mathop {\lim }\limits_{\lambda {\omega _R} \to 2} y(t) = \frac{1}{{{M_R}}}\left[ {1 - {\omega _R}t} \right]{e^{ - {\omega _R}t}}
\end{equation}\\

The six basic response functions of the three-parameter inertoviscoelastic solid A shown in Fig. 3 (top) are summarized in Table 2. 
\begin{table}[t]
\caption{Basic Frequency-Response Functions and Their Corresponding Causal Time-Response Functions of The Three-Parameter Inertoviscoelastic Solid A and B.}  
\centering
	 \includegraphics[width=.98\textwidth]{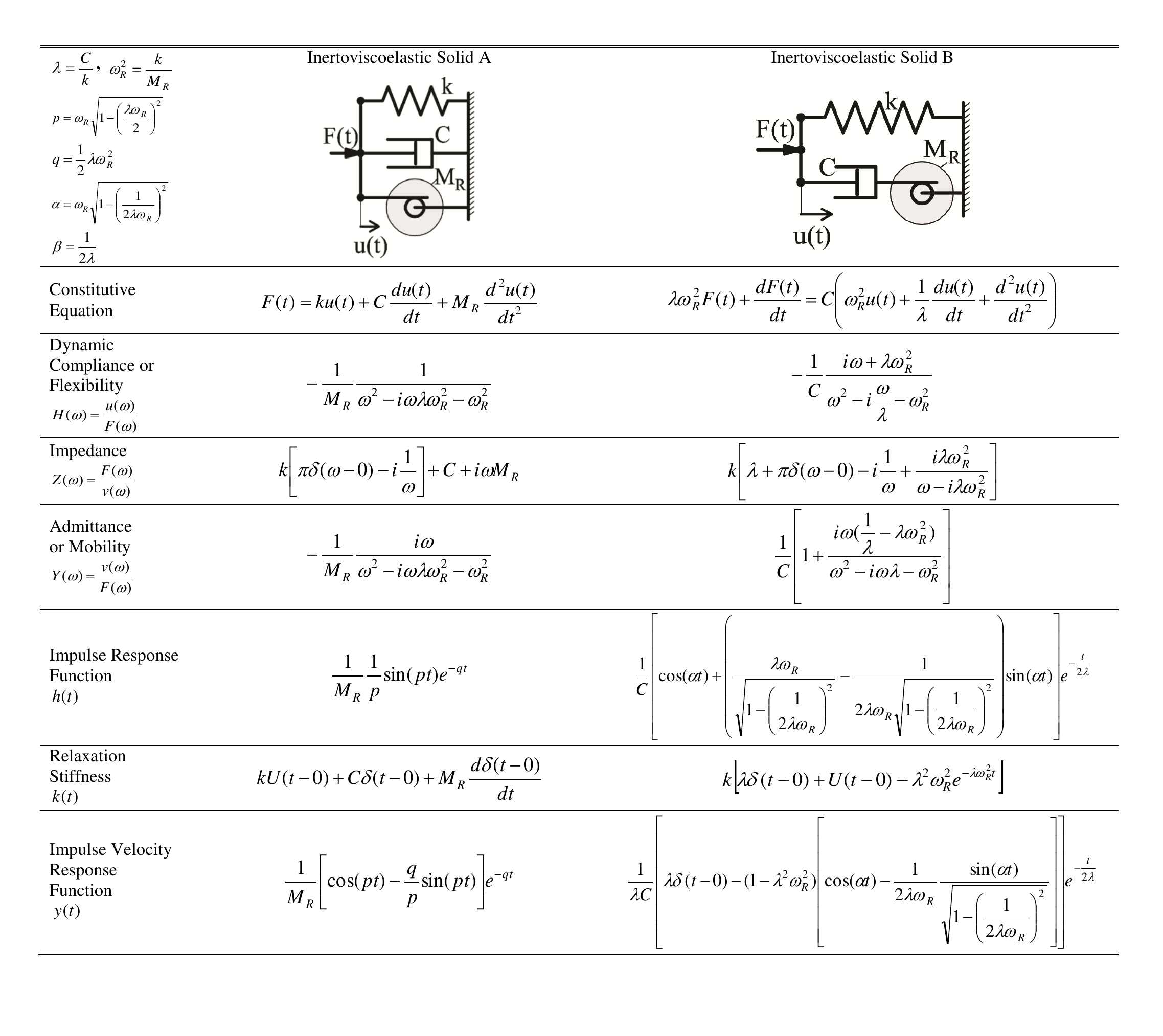}

\label{tab:2}  

\end{table}

\section*{Frequency- and Time-Response Functions of The Three-Parameter Inertoviscoelastic Solid B}
\label{sec9}

The inertoviscoelastic solid B shown in Fig. 3 (bottom) is a dashpot-inerter in-series connection that is connected in parallel with an elastic spring and has been also proposed as a conceptual vibration control element for buildings \cite{9}. By adding the forces resulting from the two parallel elements, the constitutive equation of the three-parameter mechanical network shown in Fig. 3 (bottom) is

\begin{equation}
\lambda \omega _R^2F(t) + \frac{{dF(t)}}{{dt}} = C\left( {\omega _R^2u(t) + \frac{1}{\lambda }\frac{{du(t)}}{{dt}} + \frac{{{d^2}u(t)}}{{d{t^2}}}} \right)
\end{equation}\\
where again $\lambda=C/k$ is the relaxation time and ${\omega _R} = \sqrt {k/{M_R}} $ is the rotational frequency of the network.

The Fourier transform of the constitutive equation of the inertoviscoelastic solid B given by Eq. (97) gives

\begin{equation}
F(\omega )\left( {\lambda \omega _R^2 + i\omega } \right) = C\left( {\omega _R^2 + i\frac{\omega }{\lambda } - {\omega ^2}} \right)u(\omega )
\end{equation}\\

Its dynamic compliance, $H(\omega)$ as defined by Eq. (5) is

\begin{equation}
H(\omega ) = \frac{{u(\omega )}}{{F(\omega )}} =  - \frac{1}{C}\frac{{\lambda \omega _R^2 + i\omega }}{{{\omega ^2} - i\frac{\omega }{\lambda } - \omega _R^2}}
\end{equation}\\
which is a strictly proper transfer function. The impulse response function of the inertoviscoelastic solid B is the inverse Fourier transform of the dynamic compliance given by Eq. (99)

\begin{equation}
h(t) =  - \frac{1}{{2\pi }}\frac{1}{C}\int\limits_{ - \infty }^\infty  {\frac{{i\omega  + \lambda \omega _R^2}}{{(\omega  - {\omega _1})(\omega  - {\omega _2})}}{e^{i\omega t}}d\omega } 
\end{equation}\\
where $\omega_1$, $\omega_2$ are the poles of the rhs of Eq. (99):

\begin{equation}
\begin{array}{l}
{\omega _1} = {\omega _R}\sqrt {1 - {{\left( {\frac{1}{{2\lambda {\omega _R}}}} \right)}^2}}  + i\frac{1}{{2\lambda }} = \alpha  + i\beta 
\\
\\
{\omega _2} =  - {\omega _R}\sqrt {1 - {{\left( {\frac{1}{{2\lambda {\omega _R}}}} \right)}^2}}  + i\frac{1}{{2\lambda }} =  - \alpha  + i\beta 
\end{array}
\end{equation}\\

The inverse Fourier transform of the rhs of Eq. (101) is evaluated with the method of residues and the impulse response function of the three-parameter mechanical network shown in Fig. 3 (bottom) is

\begin{equation}
\begin{array}{l}
h(t) = \frac{1}{C}[\cos (\alpha t) + (\frac{{\lambda {\omega _R}}}{{\sqrt {1 - {{\left( {\frac{1}{{2\lambda {\omega _R}}}} \right)}^2}} }}\\
{\rm{         }} - \frac{1}{{2\lambda {\omega _R}\sqrt {1 - {{\left( {\frac{1}{{2\lambda {\omega _R}}}} \right)}^2}} }})\sin (\alpha t)]{e^{ - \frac{t}{{2\lambda }}}}
\end{array}
\end{equation}\\

In the limiting case where the spring in the inertoviscoelastic solid B vanishes, $k=1/\lambda=\omega_R=0$, equation (102) reduces to Eq. (41) which is the impulse response function of a dashpot and an inerter connected in-series.

When the dimensionless quantity, $\lambda \omega_R=1/2$, then $\alpha=0$, $\beta=\omega_R$ and the three-parameter network shown in Fig. 3 (bottom) becomes critically damped. In this case ($\lambda \omega_R=1/2$), Eq. (102) assumes the expression

\begin{equation}
\mathop {\lim }\limits_{\lambda {\omega _R} \to 1/2} h(t) = \frac{1}{C}\left[ {1 - \frac{1}{2}{\omega _R}t} \right]{e^{ - {\omega _R}t}}
\end{equation}\\

Fig. 7 plots the time history of the normalized impulse response functions, $C.h(t)$, of the inertoviscoelastic solid B for values of $\lambda \omega_R=0.5$, $1$, $1.5$ and $2$.
\begin{figure}[t]
\centering
  \includegraphics[scale=0.35]{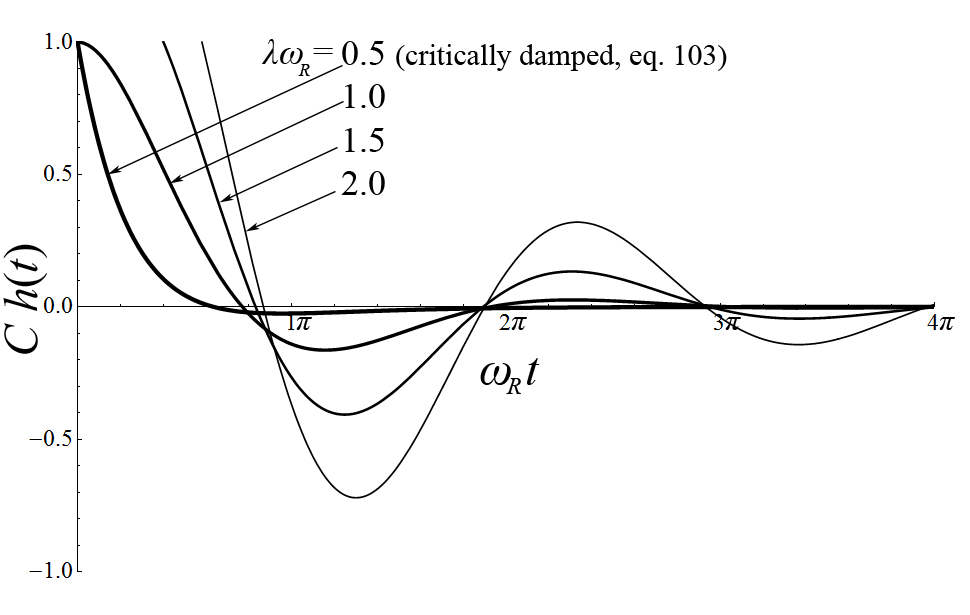}
\caption{Normalized impulse response function $C.h(t)$ of the inertoviscoelastic solid B.}
\label{fig:7}
\end{figure}

The impedance of the inertoviscoelastic solid B derives directly by Eq. (98) by using that $v(\omega)=i\omega u(\omega)$. Upon separating the high-frequency limit, $C$, the impedance of the three-parameter mechanical network shown in Fig. 3 (bottom) is expressed as

\begin{equation}
Z(\omega ) = \frac{{F(\omega )}}{{v(\omega )}} = C\left[ {1 + \frac{{\omega _R^2 + i\omega (\frac{1}{\lambda } - \lambda \omega _R^2)}}{{i\omega (\lambda \omega _R^2 + i\omega )}}} \right]
\end{equation}\\
where the frequency dependent term in the rhs of Eq. (104) is as strictly proper transfer function which has poles at $\omega=0$ and $\omega=i\lambda \omega_R^2$. Accordingly, partial fraction expansion of the frequency-dependent term gives

\begin{equation}
Z(\omega ) = k\left[ {\lambda  - \frac{i}{\omega } - \frac{{{\lambda ^2}\omega _R^2}}{{\lambda \omega _R^2 + i\omega }}} \right]
\end{equation}\\

The second term in the rhs of Eq. (105) is the impedance of the solitary spring. Accordingly, the singularity, $-i/\omega$, is enhanced with its real Hilbert companion, $\pi \delta(\omega-0)$, as indicated by Eq. (30); and the correct expression for the impedance of the inertoviscoelastic solid B is

\begin{equation}
Z(\omega ) = k\left[ {\lambda  + \pi \delta (\omega  - 0) - i\frac{1}{\omega } + \frac{{i\lambda \omega _R^2}}{{\omega  - i\lambda \omega _R^2}}} \right]
\end{equation}\\

The inverse Fourier transform of the impedance as expressed by Eq. (106) gives the causal relaxation stiffness of the inertoviscoelastic solid B

\begin{equation}
k(t) = k\left[ {\lambda \delta (t - 0) + U(t - 0) - {\lambda ^2}\omega _R^2{e^{ - \lambda \omega _R^2t}}} \right]
\end{equation}\\

In the limiting case where the spring in the inertoviscoelastic solid B vanishes, $k=1/\lambda=0$, Eq. (107) reduces to Eq. (49).

The admittance (mobility) of the inertoviscoelastic solid B shown in Fig. 3 (bottom) is the inverse of the impedance as expressed by Eq. (104). By separating its high-frequency limit, $1/C$, the admittance is expressed as

\begin{equation}
Y(\omega ) = \frac{1}{C}\left[ {1 + \frac{{i\omega (\frac{1}{\lambda } - \lambda \omega _R^2)}}{{{\omega ^2} - i\omega \lambda  - \omega _R^2}}} \right]
\end{equation}\\

The frequency-dependent term within the brackets of Eq. (108) is a strictly proper transfer function and its poles are given by Eq. (101). The inverse Fourier transform of the admittance function given by Eq. (108) is the impulse velocity response function of the three-parameter mechanical network shown in Fig. 3 (bottom) and is evaluated with the method of residues

\begin{equation}
\begin{array}{l}
y(t) = \frac{1}{{\lambda C}}[\lambda \delta (t - 0) - \\
{\rm{          }}(1 - {\lambda ^2}\omega _R^2)[\cos (\alpha t) - \frac{1}{{2\lambda {\omega _R}}}\frac{{\sin (\alpha t)}}{{\sqrt {1 - {{\left( {\frac{1}{{2\lambda {\omega _R}}}} \right)}^2}} }}]]{e^{ - \frac{t}{{2\lambda }}}}
\end{array}
\end{equation}\\
where again, $\alpha  = {\omega _R}\sqrt {1 - {{\left( {\frac{1}{{2\lambda {\omega _R}}}} \right)}^2}} $. In the limiting case where the spring in the inertoviscoelastic solid B vanishes, $k=1/\lambda=\omega_R=0$ and Eq. (109) reduces to Eq. (55).

When the dimensionless quantity, $\lambda \omega_R=1/2$, then $\alpha=0$, $\beta=\omega_R$ and the three-parameter network shown in Fig. 3 (bottom) becomes critically damped. In this case ($\lambda \omega_R=1/2$), Eq. (109) assumes the expression

\begin{equation}
\mathop {\lim }\limits_{\lambda {\omega _R} \to 1/2} y(t) = \frac{1}{{\lambda C}}\left[ {\lambda \delta (t - 0) - \frac{3}{4}(1 - {\omega _R}t){e^{ - {\omega _R}t}}} \right]
\end{equation}\\

The six basic response functions of the three-parameter inertoviscoelastic solid B shown in Fig. 3 (bottom) are summarized in Table 2 next to the basic response function of the inertoviscoelastic solid A.

\section*{Conclusions}
\label{sec10}

This paper derives the causal time-response functions of three-parameter mechanical networks which involve the inerter--a two-node element in which the force-output is proportional to the relative acceleration of its end-nodes. This is achieved by extending the relation between the causality of a time-response function with the analyticity of its corresponding frequency response function to the case of generalized functions.\\

The paper shows that when the frequency-response function has as singularity the reciprocal function, $1/(\omega-\omega_R)$ (with $\omega_R$=constant or zero), the complex frequency-response function needs to be enhanced with the addition of a Dirac delta function, $\delta(\omega-\omega_R)$, so that the real and imaginary parts of the correct frequency-response function are Hilbert pairs; therefore, yielding a causal time-response function in the time domain. Similarly, when the singularity of the frequency-response function is $1/\omega^2$, the complex frequency-response function needs to be enhanced with the addition of $d \delta(\omega-0)/d\omega$.\\

Table 1 confirms that the basic response functions of mechanical networks which involve inerters follow the same superposition rules observed in the basic response functions of classical mechanical networks that involve just springs and dashpots. For instance, the dynamic compliance (flexibility) and admittance (mobility) of the three-parameter inertoviscoelastic fluids A, B and C (shown in Fig. 2) are the summation of the corresponding compliances or admittances of the solitary element (inerter in fluid A, spring in fluid B and dashpot in fluid C) with the compliances or admittances of the two-element in-parallel connection network (Kelvin-Voight model in A, rotational inertia damper in B and inertoelastic solid in C). The outcome of this superposition is reflected in the resulting causal time-response functions which are the impulse response function, $h(t)$, and the impulse velocity response function, $y(t)$. Similar superposition rules shown in Table 2 apply in the impedances $Z(\omega)=F(\omega)/v(\omega)$ and the corresponding causal relaxation stiffness, $k(t)$, of the inertoviscoelastic solids A and B shown in Fig. 3.\\

The integral representation of the output signals presented in this study offers an attractive computational alternative given that the constitutive equations of some of the three-parameter models examined involve the third derivative of the nodal displacement (derivative of the acceleration) which may challenge the numerical accuracy of a state-space formulation given that in several occasions the input signal is only available in digital form as in the case of recorded accelerograms.

\end{document}